\documentclass[twocolumn,showpacs,showkeys,preprintnumbers,amsmath,amssymb]{revtex4}  
 
\usepackage{graphicx}
\usepackage{dcolumn}
\usepackage{bm}
\usepackage{siunitx}

\begin{document}


\title{Voltage control of electron-nuclear spin correlation time in a single quantum dot}

\author{ J. Nilsson$^{1}$, L. Bouet$^{2}$, A.J. Bennett$^{1}$, T. Amand$^{2}$, R.M. Stevenson$^{1}$, I. Farrer$^{3}$, D.A. Ritchie$^{3}$, S. Kunz$^{2}$, X. Marie$^{2}$,A.J. Shields$^{1}$,B. Urbaszek$^{2}$}

 \affiliation{%
$^1$Cambridge Research Laboratory, Toshiba Research Europe Limited, 208 Science Park, Milton Road, Cambridge, CB4 0GZ, United Kingdom}

\affiliation{%
$^2$Universit\'e de Toulouse, INSA-CNRS-UPS, LPCNO, 135 Av. Rangueil, 31077 Toulouse, France}

\affiliation{%
$^3$Cavendish Laboratory, University of Cambridge, JJ Thomson Avenue, Cambridge, CB3 0HE, United Kingdom}

\date{\today}

\begin{abstract}
We demonstrate bias control of the hyperfine coupling between a single electron in an InAs quantum dot and the surrounding nuclear spins monitored through the positively charged exciton X$^+$ emission. In applied longitudinal magnetic fields we vary simultaneously the correlation time of the hyperfine interaction and the nuclear spin relaxation time and thereby the amplitude of the achieved dynamic nuclear polarization under optical pumping conditions. In applied transverse magnetic fields a change in the applied bias allows to switch from the anomalous Hanle effect to the standard electron spin depolarization curves. 

\end{abstract}

\pacs{72.25.Fe,73.21.La,78.55.Cr,78.67.Hc}
                            \keywords{Quantum dots, hyperfine interaction}
\maketitle

\begin{figure}
\includegraphics[width=0.38\textwidth]{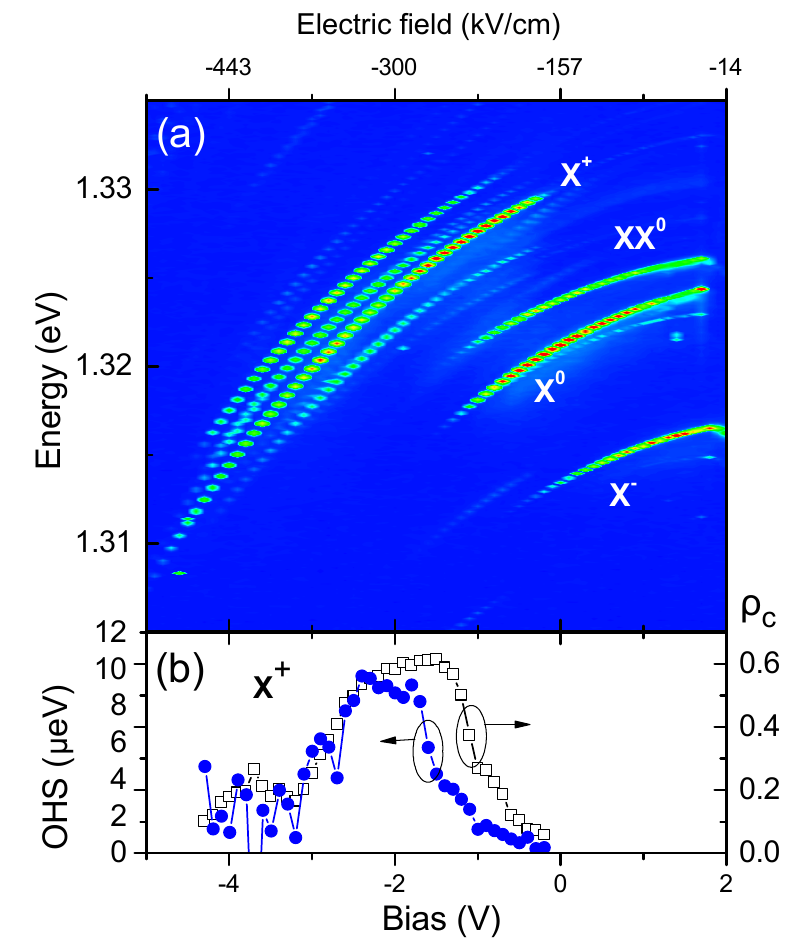}
\caption{\label{fig:zerofield} Dot A. (a) Contour map of PL versus applied voltage in zero external magnetic field B=0. Around -1.3~V the electron tunnelling rate out of the dot has been sufficiently suppressed to permit the existence of neutral exciton (X$^0$) and biexciton (XX$^0$) complexes and at $\sim$0~V the negative trion (X$^-$) appears as electrons tunnel into the dot at a faster rate (b) Overhauser shift OHS (closed circles) and degree of circular polarisation $\rho_c$ (open squares) for the X$^{+}$ transition at B=0.}
\end{figure}

\textit{Introduction.---}Localised carriers in semiconductors strongly interact with the nuclear spins of the lattice atoms via the hyperfine interaction \cite{Meier:1984a}. The main applications for self assembled semiconductor quantum dots (QDs) for quantum information processing demand well defined carriers spin states with long coherence times. Controlling the hyperfine interaction is a vital prerequisite for spin-manipulation schemes [\onlinecite{Bluhm:2010b, Boyer:2010}], entangled photon emitters \cite{Stevenson:2011a,Kuroda:2013a} and spin-photon entanglement \cite{Degreve:2012a,Gao:2012a,Schaibley:2013a}. Manipulating the nuclear spin system is interesting in its own right due to the non-linear (quantum) dynamics observed \cite{Urbaszek:2013a} and the prospects  of nuclear spintronics \cite{Reimer:2010a}. The latter aims to benefit from the robustness of the nuclear spin system but future implementations still require more practical, complementary control schemes to RF pulses to manipulate the nuclear spins. Here we aim to explore the powerful combination of optical and electric field control of the mesoscopic nuclear spin system in a single quantum dot at T=4K, providing a complementary approach to the transport studies in gate defined quantum dots at mK temperatures \cite{Bluhm:2010b,Reilly:2008a}. \\
\indent The achievable degree of nuclear spin polarization depends on the strength of the hyperfine interaction. Taking into account the fixed carrier confinement given by the QD shape and size the main parameter that can be varied experimentally is the electron correlation time of the hyperfine interaction $\tau_c$. A finite $\tau_c$ introduces a broadening (overlap) of the electron Zeeman levels that allows for spin flips with nuclei to take place without violating energy conservation. We demonstrate tuning with bias of $\tau_c$ by a factor of $\geq 6$ for InGaAs QDs embedded in a diode structure and observe a drastic increase in the nuclear spin relaxation time as $\tau_c$ is decreased. The structures allow to cover an applied electric field range of $\sim$200 kV/cm without adding extra charges to the dot [\onlinecite{bennett_nphys_2010}], far wider than what has been possible in previous studies of dynamic nuclear polarization (DNP) with charge-tunable structures \cite{Makhonin:2009,MaletinskyPRB2007}. This enables us to control the width of the bistablity region of both the electron and nuclear spin system in applied longitudinal magnetic fields. In transverse magnetic fields $B_x$ (i.e. perpendicular to the growth direction and light propagation axis), we observe a drastic departure from the standard Hanle depolarization curve for electrons \cite{Krebs:2010a}. The electron precession around the applied field is clearly inhibited for $B_x$ as high as 0.9T. By varying the applied bias we show the gradual transition from this anomalous to the standard Hanle effect for the same QD as the hyperfine interaction strength is tuned. Our results obtained with non-resonant excitation in the wetting layer show that the quasi-resonant excitation used in the initial report of the anomalous Hanle effect \cite{Krebs:2010a} is not a key ingredient for the build-up of the transverse nuclear spin polarization that stabilizes the electron.\\
\indent \textit{Sample and experimental methods.---}Our sample contains self-assembled InAs QDs embedded in a GaAs diode structure. The design includes thick AlGaAs barriers to allow for wide range electric field tuning while avoiding the tunnelling out of photo-generated carriers. Such devices have previously been used to tune the energy of different exciton complexes [\onlinecite{bennett_apl_2010}], to observe the coherent coupling between neutral exciton states [\onlinecite{bennett_nphys_2010}] and to control the electron- and hole g-factors [\onlinecite{bennett2013}]. Distributed Bragg reflectors placed outside the tunnelling barriers create an optical cavity (weak coupling), enhancing the light collection efficiency around the emission wavelengths of the quantum dots ($\sim$940 nm). 
We excite directly the heavy hole to electron transition at 1.43 eV in the wetting layer with a cw Ti-Sa laser \cite{BelhadjPRL}. The photoluminescence (PL) spectra at T=4K are recorded using a spectrometer equipped with a Si-CCD. A combination of liquid crystal retarders and Glan-Taylor polarisers allow us to achieve circular ($\sigma^+$, $\sigma^-$), linear ($X$, $Y$) or arbitrary elliptical light excitation and detection.\\
\indent \textit{Dynamic nuclear polarisation in zero external magnetic field.---}
At the centre of this study is the X$^+$ transition (2 holes in a spin singlet, 1 electron), which is stable over an electric field range of  $\sim$200 kV/cm, see Fig. \ref{fig:zerofield}(a). During the radiative lifetime $\tau_r$ of the X$^+$, the mean spin  $\langle  S_z \rangle$ of the unpaired electron is partially transferred to the nuclear spin system via the Overhauser effect \cite{Overhauser:1953a,hole}. This results in the build-up of an effective nuclear magnetic field $B_{n,z}$ that will in turn act on the electron. The Overhauser shift (OHS), $\Delta E_{OHS} = |g_e \mu_B B_{n,z}|$ is detected as the energy difference between $\sigma^+$ and $\sigma^-$ polarised emission lines. The degree of circular polarisation $\rho_c = (I_{\sigma^+} -  I_{\sigma^-})/(I_{\sigma^+} + I_{\sigma^-})= -2 \langle  S_z \rangle$  allows to probe directly the electron spin polarisation. Both OHS and $\rho_c$ are shown in figure \ref{fig:zerofield}(b) for X$^{+}$. Note that no external magnetic field is applied here. Around -2V, roughly in the middle of the X$^{+}$ range, an OHS of $\sim 8\mu$eV and a high degree of PL circular polarisation ($\sim$60\%)  are observed. Here, $B_{n,z}$ helps to stabilise the electron spin against the nuclear field fluctuations $\delta B_n$ in the QD [\onlinecite{BelhadjPRL}]. 

\begin{figure*}[!]
\includegraphics[width=0.85\textwidth]{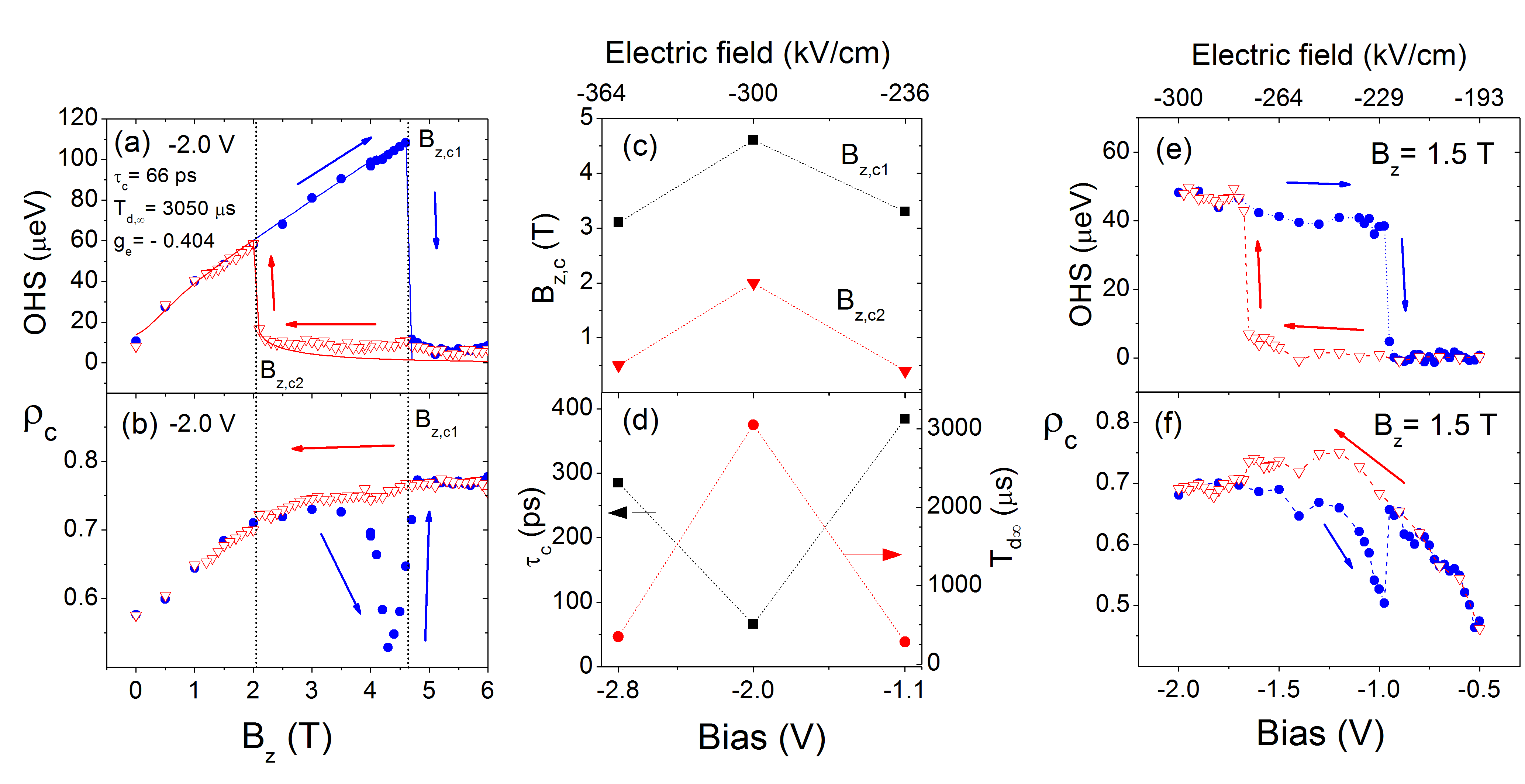}
\caption{\label{fig:dnp_faraday} Dot A. (a) Overhauser shift OHS and fit with Eq.(1) (solid lines) (b) degree of circular polarisation $\rho_c$ of the X$^+$ transition in magnetic field ($B_z$) sweep for constant bias -2.0 V. Closed circles: sweeping the applied magnetic field up, open triangles sweeping down. (c) Critical fields for magnetic field sweep up ($B_{z,1}$) and down ($B_{z,2}$) for different bias voltages. (d) Bias dependence of electron hyperfine correlation time ($\tau_c$) and nuclear depolarisation time ($T_{d,\infty}$) determined with an accuracy of 10\%. (e) OHS bistability when sweeping the applied voltage first in positive, then in negative direction, for constant $B_z = 1.5$~T. (f) Corresponding $\rho_c$. Closed circles: sweeping applied voltage in positive direction, open triangles negative direction. Dotted lines: guide to the eye}
\end{figure*}

\textit{Faraday Geometry: Voltage control of dynamic nuclear polarization.---} The results presented in Fig. \ref{fig:zerofield} on their own strongly indicate that we can control the electron-nuclear interaction by changing the bias voltage. To study the voltage dependence further, we perform DNP measurements in applied longitudinal magnetic fields $B_z$ at different applied voltages. For each voltage $B_z$ is swept first from zero to a maximum of 6T, and back down to zero, all the time exciting spin-polarised electrons with a $\sigma^{+}$-polarised laser and recording the PL-spectra. Figure \ref{fig:dnp_faraday}(a) shows the measured OHS at constant bias -2.0~V. As the external field is ramped up the OHS increases and compensates the electron Zeeman energy, keeping the total electron spin splitting $\Delta E_e = g_e \mu_B \left( B_z + B_{n,z}\right)$ close to zero. This ensures that electron-nuclear spin flip-flops and hence further build-up of nuclear spin polarisation remain possible. At the critical field $B_{z,c1}=4.6$~T the OHS saturates:  the nuclear spin system is unable to compensate the external field and a sudden collapse of OHS occurs. As the external magnetic field is swept back down the electron Zeeman splitting due to the external field $B_z$ gradually decreases. An OHS around zero is observed until a critical field $B_{z,c2}=2.0$~T where the Zeeman energy is small enough for flip-flops with nuclei to take place and the OHS recovers.   
For the X$^+$ PL $\rho_c$ is proportional to the electron spin \textit{after} the interaction with the nuclei during $\tau_r$. The nuclear spins are polarized via flip-flops with the electron spins, so if this spin flip rate becomes comparable to $1/\tau_r$ the measured $\rho_c$ should decrease. This explains the drastic decrease in $\rho_c$ in Fig. \ref{fig:dnp_faraday}(b) for $B_z$ just below 4.5T  as the system aims to maintain $|B_{n,z}|\approx |B_z|$\cite{comp,Urbaszek:2013a}.

\begin{figure}[!]
\includegraphics[width=0.5\textwidth]{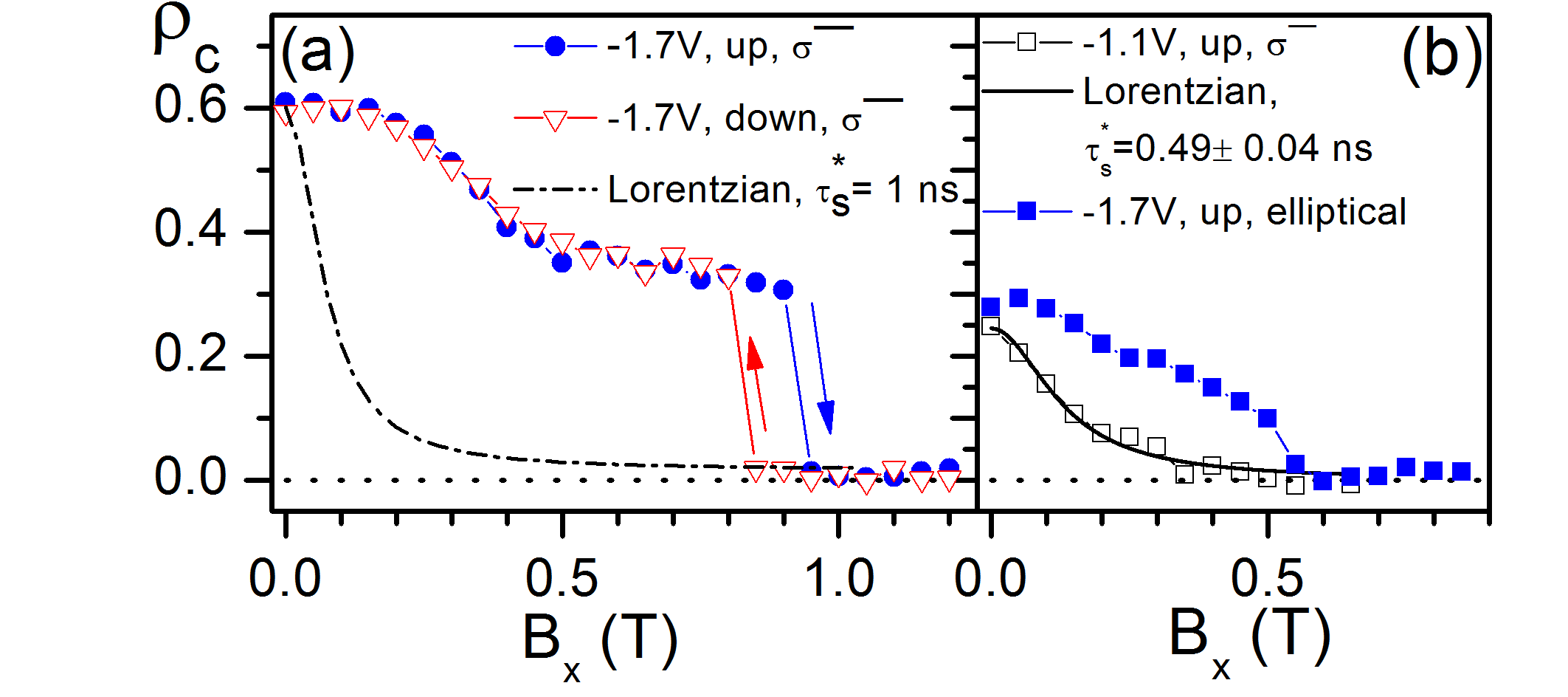}
\caption{\label{fig:qda_hanle} Dot A. (a) Hanle measurements at -1.7V using $\sigma^-$ excitation sweeping the transverse field $B_x$ up (blue full circles) and down (red open triangles).(b) Hanle measurements at bias -1.1V (open squares), together with measurements carried out at -1.7V with elliptically polarised excitation (full squares). A Lorentzian fit (black line) to the -1.1V data allows to extract $\tau_s^*$}
\end{figure}

At two different bias voltages, -1.1~V and -2.8~V, we also observe OHS hysteresis, albeit with very different critical magnetic field values, as can be seen in Fig. \ref{fig:dnp_faraday}(c). To quantify the dependence of the hyperfine correlation time $\tau_c$ and the nuclear spin relaxation time $T_d$ on the applied bias, we fit the OHS dependence on $B_z$ for the experiments carried out at different bias values with an implicit equation [\onlinecite{Eble2006}]:
%
%
%
\begin{equation}
{
 \Delta E_{OHS} = g_e \mu_B B_{n,z} = - \frac{2\tilde{A}\tilde{Q}~( \langle S_z \rangle - \langle S_z \rangle _0  )}{ 1 + T_{1e}(B_{n,z})/T_d(B_{n,z}) } 
}
\label{eq:impliciteq}
\end{equation}
where $T_{1e}^{-1} = (\tilde{A}/N\hbar)^2 \cdot 2f_e \tau_c /[1 + (\Delta E_e(B_{n,z}) \tau_c / \hbar)^2]$ is the electron-induced spin-flip rate. The nuclear spin relaxation time $T_d$ through e.g. nuclear dipole-dipole interactions is modelled with a phenomenological magnetic field dependence capturing its inhibition in strong magnetic fields [\onlinecite{Krebs2008}]: $T_d^{-1} =  T^{-1}_{d\infty} + T^{-1}_{d0}/[1 + (B_z/B_Q)^2] $ . 
This model has previously found application in explaining the evolution of OHS with excitation polarization [\onlinecite{UrbaszekTemp2007,Braun:2006a}] and magnetic field [\onlinecite{MaletinskyPRB2007,KajiAPL2007}]. \\
\indent To fit the hysteresis curves as in Fig.\ref{fig:dnp_faraday}(a) for all three bias values, the following parameters are kept constant: $\tilde{A} \approx 47\mu$eV as the average hyperfine constant for InAs quantum dots, $\tilde{Q} = 13$, the average electron spin created at the start of the interaction between the electron and nuclei $\langle S_z \rangle$ is taken as a first approximation to be constant $\langle S_z \rangle \sim 0.4$ \cite{szero}, $B_Q =0.4$~T, $N = 6 \cdot 10^4$, $T_{d0} = $\SI{500}{\micro s} \cite{Urbaszek:2013a}.\\
\indent Only three parameters in Eq.\ref{eq:impliciteq} had to be varied as the bias changed. 
The bias dependence of the longitudinal g factor $g_e$ variation has already been determined in the same structure \cite{bennett2013}.
The two most sensitive parameters are the electron correlation time $\tau_c$ and the nuclear depolarisation time $T_{d\infty}$, determined primarily by the critical fields $B_{z,c1}$ and $B_{z,c2}$ that change clearly with bias. Figure \ref{fig:dnp_faraday}(a) includes the resulting fit to the OHS at -2.0V, which shows excellent quantitative agreement. Figure \ref{fig:dnp_faraday}(d) shows the extracted values for $\tau_c$ and $T_{d\infty}$ as a function of applied bias voltage. At -2.0V, where the largest OHS can be built up, we find a minimum for $\tau_c$ and a maximum for $T_{d\infty}$. Away from -2.0~V $\tau_c$ increases while $T_{d\infty}$ decreases. Such reciprocal behaviour of $\tau_c$ and $T_{d\infty}$ has been observed before \cite{UrbaszekTemp2007}. Our results strongly suggest that the presence of a spin polarized electron in the dot, given by $\tau_c$, has a direct influence on the nuclear spin relaxation time $T_d$. This agrees with the observation of long nuclear spin memory times only for unpopulated QDs \cite{Maletinsky:2009}. Nuclear spin depolarisation mechanisms linked to the presence of the electron in the dot could be direct through Knight field fluctuations \cite{Huang:2010a}, or indirect through local fluctuations of the electric field gradients, i.e. the nuclear quadrupole effects, induced by the presence of a charge carrier \cite{Paget:2008a}. So shortening $\tau_c$ by applying a bias of -2.0V has two positive effects if we aim for robust nuclear spin polarization: First, a short $\tau_c$ gives a large uncertainty in the electron Zeeman energy which is favorable for spin flip-flops. Second, a QD being populated only during a short time $\tau_c$ by an electron will keeps its nuclear spin polarization for longer.\\
\indent A very direct way to probe the voltage control of the OHS, enabled by the wide tuning range of our device, is to scan the bias while recording the X$^+$ PL \cite{Makhonin:2009}. Figure \ref{fig:dnp_faraday}(e) shows a bias scan of the measured OHS for constant external field $B_z=1.5$ T, and \ref{fig:dnp_faraday}(f) shows the corresponding $\rho_c$. A bistability for both nuclear and electron spins is observed, with a collapse at $-1.0$~V in a bias region with longer hyperfine correlation time, and a recovery when the voltage is $-1.7$~V with a shorter correlation time (compare with Fig. \ref{fig:dnp_faraday}(d)). \\
\indent \textit{Voigt Geometry: Voltage control of anomalous Hanle effect.---} Nuclear spin effects have been proposed to stabilize electrons spins against depolarization in applied transverse magnetic fields \cite{Krebs:2010a,Dzhioev:2007a}. Here we use bias control to tune the nuclear spin polarization with drastic consequences for the electron spin system, see Fig. \ref{fig:qda_hanle}. The experiments start at zero applied field by constructing DNP under optical pumping of the wetting layer. Then we gradually increase the amplitude of an applied \textit{transverse} magnetic field $B_x$ and monitor the electron spin depolarization via the X$^+$ PL emission from the QD. In a simple picture, the initialized electron spin as well as the nuclear spins precess around the applied field and a standard Hanle depolarization is expected \cite{Meier:1984a}. Previous studies of the X$^+$ PL polarization in single InGaAs dots have revealed a clear departure from this simple scenario \cite{Krebs:2010a} as the nuclear spin system is subject to strong quadrupole effects equivalent to internal fields in the hundreds of mT range \cite{Bulutay:2012a}. Hanle measurements on ensembles of InGaAs QDs \cite{Verbin:2012a} analysing the amplitude of the negative PL polarization of the X$^-$ have shown depolarization curves similar to localised electrons in bulk GaAs\cite{Paget:1977a}. The reported Hanle curves are far narrower than reported by \textcite{Krebs:2010a} for single dots. Here we show that depending on the applied bias and the optical pumping conditions both types of results can be obtained for the same QD \cite{wshape}. 

For a constant bias of -1.7~V both the electron and nuclear spin polarization are maximised (Fig.\ref{fig:zerofield}(b)). We measure at $B_x=0$ for X$^+$ PL $\rho_c=60\%$ which remains constant up to $B_x=0.2$~T, see Fig.\ref{fig:qda_hanle}(a). While increasing the magnetic field further, the polarization decreases only gradually, with a strong polarization of $\rho_c = 35\%$ remaining at 0.9~T. For the next $B_x$ value at 0.95~T the electron spin polarization has collapsed as $\rho_c=0$. In the same experimental run we now gradually decrease $B_x$ and for $B_x = 0.8$~T $\rho_c$ recovers  to a value of 35\%. We observe a clear bistability region of the electron spin polarization as a function of the applied transverse magnetic field $B_x$ \cite{voigt}. Also shown in Fig.\ref{fig:qda_hanle}(a) is the theoretically expected Lorentzian depolarisation curve, emphasizing the radical deviation of our observations from the standard Hanle effect. At the origin of the anomalous Hanle effect the build-up of a transverse nuclear magnetic field $B_{n,x}$ has been proposed and observed \cite{Krebs:2010a}, that practically cancels the applied field $B_x$. As a result the electron is subject to a total field that is close to zero and it keeps its initial spin orientation. How an internal nuclear field \textit{orthogonal} to the initial electron spin orientation can be constructed in these strained dots still needs to be understood. In this geometry recently evoked  non-collinear terms of the hyperfine interaction that do not require a carrier spin flip to construct DNP should become important  \cite{Hogele:2012a,Yang:2012a}. In our experiments spin polarized electrons are generated in the wetting layer before being captured (one at a time) by the dot, which is in contrast to the initial report of the anomalous Hanle effect achieved under quasi-resonant excitation. This shows that the anomalous Hanle effect is robust and relies only on the generation of a mean electron spin $\langle S_z \rangle \neq 0$ without any particular phase.\\
\indent Independent of the exact nuclear spin polarization mechanism, changing the applied bias is expected to result also in Voigt geometry in a strong variation of the correlations time of the hyperfine interaction and possibly in a modulation of the local electric field gradients i.e. of the nuclear quadrupole effects. At -1.1~V we would expect hyperfine coupling to have less impact as the correlation time increases (Fig. \ref{fig:dnp_faraday}(d)). The depolarization curve at -1.1V shown in Fig. \ref{fig:qda_hanle}(b) does indeed show a stark difference compared to data taken at -1.7~V, going smoothly to zero as $B_x$ is increased. The electron depolarization follows a Lorentzian curve \cite{Meier:1984a}, allowing us to extract the spin lifetime $\tau_s^*$ as $B_{1/2}=\hbar/(|g_{e,\bot}|\mu_B \tau_s^*)$. Using the electron g-factor of 0.181 extracted for this dot, we obtain $\tau_s^*\simeq 490$~ps which is only slightly shorter than the radiative lifetime $\tau_r$. The initial polarization value $\rho_c$ for a bias of -1.1~V at $B_x=0$ was 25\%, considerably lower than the 60\% at -1.7~V. To see if the initial polarization is the main reason for the drastic change in the Hanle curve, we have performed measurements at -1.7~V under elliptical laser excitation in order to generate the exact same initial polarization values as at -1.1~V. At -1.7~V under elliptically polarized laser excitation we obtain very extended Hanle curves, very different from the measurements at -1.1~V.  As can be seen clearly in Fig. \ref{fig:qda_hanle}(b), the applied bias controls whether the system will exhibit standard or anomalous Hanle depolarisation curves, and their width. The initial electron polarization at $B_x=0$ plays a role for this evolution but is clearly not the only parameter that needs to be taken into account. In light of our measurements in longitudinal magnetic fields (Fig.~\ref{fig:dnp_faraday}) we suggest that the hyperfine electron correlation time $\tau_c$ is a key parameter to tune the extent of the anomalous Hanle effect. The exact deviation of the electron depolarization curve from a standard Lorentzian will depend on the strength of the hyperfine coupling during the experiments. The relative importance of the nuclear quadrupole effects will depend on the achieved degree of DNP \cite{quadvoigt}. \\
\indent \textit{Conclusion.---}The measurements in Faraday geometry show voltage control of the DNP bistability. Our model allows to extract the direct influence of the applied bias on the correlation time $\tau_c$ of the hyperfine interaction and the nuclear spin relaxation time $T_{d\infty}$. Although $\tau_c$ and $T_{d\infty}$ change by one order of magnitude as a function of the applied bias, their product $\tau_c.T_{d\infty}$ remains constant within a factor of 2. So for dots charged with electrons, fluctuations of the Knight field \cite{Huang:2010a} and of local electric field gradients \cite{Urbaszek:2013a} constitute a major source of nuclear spin relaxation. In applied transverse magnetic fields under non-resonant optical pumping in the wetting layer we demonstrate bias control of the amplitude and magnetic field range of the anomalous Hanle effect. In the investigated sample structure all-electrical coherent control of the exciton states on a ns time scale has been demonstrated \cite{Boyer:2010}. A natural extension of this work is to use fast electric control of the nuclear spin to gain unique insight into in the evolution of the mesoscopic nuclear spin system. \\
\indent \textit{Acknowledgments.---} We acknowledge partial support for this work from the EU through the FP7 FET Q-ESSENCE Integrated Project, ITN Spin-Optronics and ERC St. Gr. OptoDNPcontrol (B.U.).

\providecommand{\noopsort}[1]{}\providecommand{\singleletter}[1]{#1}%


\begin{thebibliography}{40}
\bibitem[{com()}]{comp}
\bibinfo{note}{At this critical magnetic field value, the electron spin might
  also be depolarized by $\delta B_n$.}
\bibitem[{sze()}]{szero}
\bibinfo{note}{The equilibrium electron spin polarisation $\langle S \rangle_0
  = 1/2 \tanh \left( g_e \mu_B B_z/k_BT\right)$ is at 4K negligible over the
  magnetic field range explored here.}
\bibitem[{voi()}]{voigt}
\bibinfo{note}{At the critical B$_x$ values where sudden jumps in $\rho_c$
  occurred we were not able to resolve any sudden change in the X$^+$
  transition energy i.e. Zeeman splitting.}
\end{thebibliography}
\end{document}